\documentclass[%
 aip,
 reprint,%
nofootinbib, %
amsmath,amssymb
]{revtex4-1}

\usepackage{graphicx}
\usepackage{dcolumn}
\usepackage{bm}

\usepackage[utf8]{inputenc}
\usepackage[T1]{fontenc}
\usepackage{mathptmx}

\usepackage{hyperref}
\hypersetup{%
	unicode=false, %
	bookmarksnumbered=true, %
	bookmarkstype=toc, %
	breaklinks=true, %
	colorlinks=true, %
	pdfborder={0 0 1}, %
	linkcolor=blue, %
	citecolor=blue, %
	urlcolor=blue, %
	pdftoolbar=true, %
	pdfmenubar=true, %
	pdffitwindow=false, %
	pdfstartview={FitH} %
}

\newcommand{\zi}{\mathrm{i}}

\newcommand{\br}{\bm{r}}
\newcommand{\bk}{\bm{k}}
\newcommand{\bp}{\bm{p}}
\newcommand{\bq}{\bm{q}}
\newcommand{\me}{m_{\mathrm{e}}}

\newcommand{\R}{\mathrm{R}}
\newcommand{\A}{\mathrm{A}}

\renewcommand{\Re}{\mathrm{Re}}
\renewcommand{\Im}{\mathrm{Im}}

\begin{document}

\title[Spin-charge conversion and current vortex]{Spin-charge conversion and current vortex in spin-orbit coupled systems}

\author{Junji Fujimoto}
\altaffiliation[Present address: ]{Department of Physics, University of Tokyo, Bunkyo, Tokyo 113-0033, Japan}
\affiliation{Kavli Institute for Theoretical Sciences, University of Chinese Academy of Sciences, Beijing, 100190, China}

\author{Florian Lange}
\affiliation{Institut f\"ur Physik, Universit\"at Greifswald, 17489 Greifswald, Germany}

\author{Satoshi Ejima}
\affiliation{Institut f\"ur Physik, Universit\"at Greifswald, 17489 Greifswald, Germany}
\affiliation{RIKEN Cluster for Pioneering Research (CPR), Wako, Saitama 351-0198, Japan}

\author{Tomonori Shirakawa}
\affiliation{RIKEN Center for Computational Science (R-CCS), Kobe, Hyogo 650-0047, Japan}

\author{Holger Fehske}
\affiliation{Institut f\"ur Physik, Universit\"at Greifswald, 17489 Greifswald, Germany}

\author{Seiji Yunoki}
\affiliation{RIKEN Cluster for Pioneering Research (CPR), Wako, Saitama 351-0198, Japan}
\affiliation{RIKEN Center for Computational Science (R-CCS), Kobe, Hyogo 650-0047, Japan}
\affiliation{RIKEN Center for Emergent Matter Science (CEMS), Wako, Saitama 351-0198, Japan}

\author{Sadamichi Maekawa}
\affiliation{RIKEN Center for Emergent Matter Science (CEMS), Wako, Saitama 351-0198, Japan}
\affiliation{Kavli Institute for Theoretical  Sciences, University of Chinese Academy of Sciences, Beijing, 100190, China}

\date{\today}

\begin{abstract}
  Using response theory, we calculate the charge-current vortex generated by spin pumping at a point-like contact in a  system with Rashba spin-orbit coupling.  
  We discuss the spatial profile of the current density for finite temperature and for the zero-temperature limit. 
  The main observation is that the Rashba spin precession leads to a charge current that oscillates as a function of the distance from the spin-pumping source, which is confirmed by numerical simulations. 
  In our calculations, we consider a Rashba model on a square lattice, for which we first review the basic properties related to charge and spin transport. 
 In particular, we define the charge- and spin-current operators for the tight-binding Hamiltonian as the currents coupled linearly with the $\mathrm{U}(1)$ and $\mathrm{SU}(2)$ gauge potentials, respectively. 
 By analogy to the continuum model, the spin-orbit-coupling Hamiltonian on the lattice is then introduced as the generator of the spin current.  
\end{abstract}

\maketitle

\section{\label{sec:intro}Introduction}
Generation and detection of non-equilibrium spin angular momentum are of crucial importance in spintronics.
In recent decades, research has been extended to the interconversion of spin and other physical quantities, such as charge~\cite{hirsch1999,murakami2003,sinova2004}, heat~\cite{uchida2008}, and mechanical angular momentum~\cite{matsuo2011,takahashi2016a,kobayashi2017}. 
The spin Hall effect (SHE) is commonly used in experiments to electrically generate spin currents, while the inverse spin Hall effect (ISHE) is used as a detector. 
The SHE and ISHE originate from spin-orbit couplings~(SOCs) through intrinsic, side-jump scattering, and skew-scattering mechanisms, which each contribute to the effects in a different way~\cite{sinova2015}.

\begin{figure}
\centering
\includegraphics[width=0.95\linewidth]{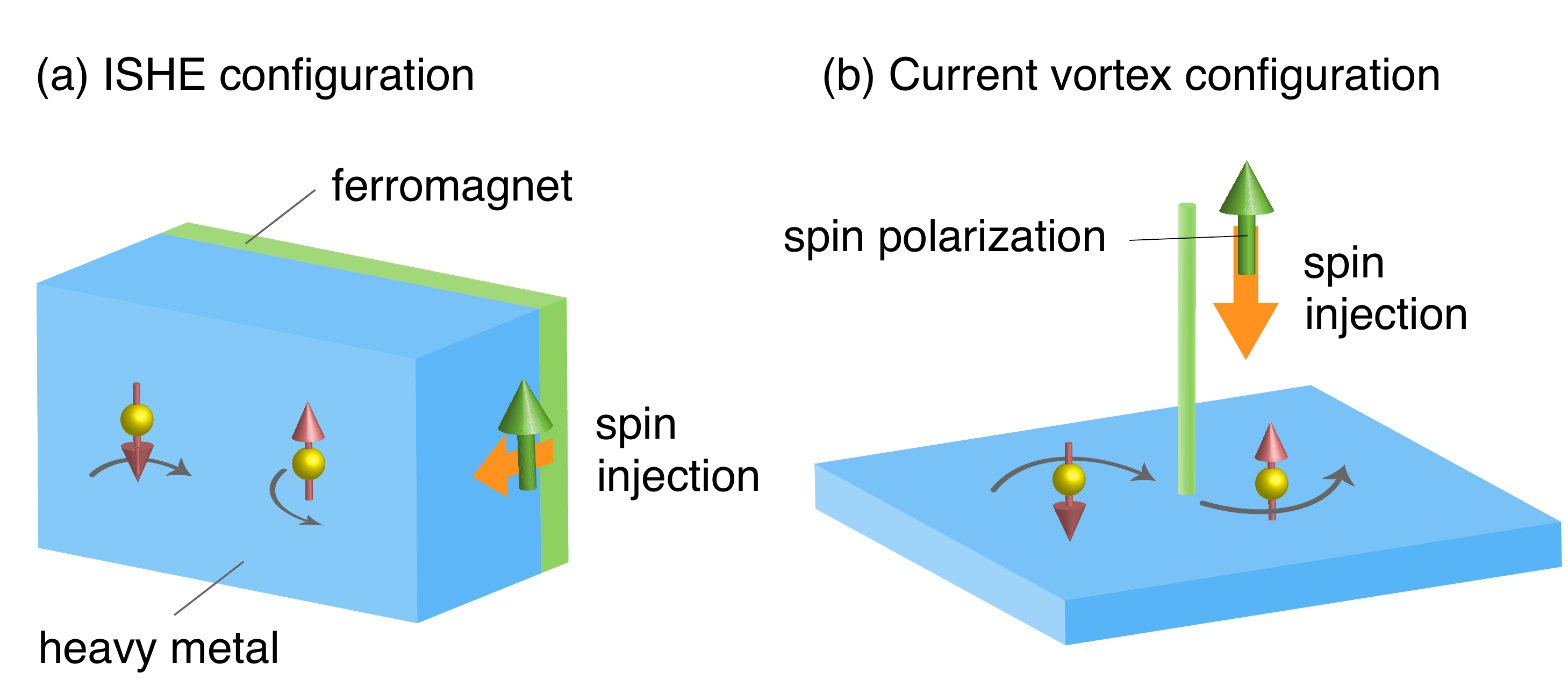}
\caption{\label{fig:1}(a)~Spin-charge conversion in a typical configuration composed of a ferromagnet and a heavy metal with strong spin-orbit coupling (SOC). The spin injection into the heavy metal from the ferromagnet by spin pumping induces an electric voltage via the inverse spin Hall effect (ISHE). %
  (b)~A different configuration, in which the spin is injected locally, inducing a nonuniform charge current. 
  For the Rashba SOC with the spin injection whose polarization is perpendicular to the plane, the charge current forms a vortex.}
\end{figure}

To experimentally gain more information about the transport behavior a system, it may be useful to consider different geometries for the spin-charge conversion besides the usual configuration for ISHE experiments~[Fig.~\ref{fig:1} (a)]. 
Here, we propose to investigate the nonuniform charge-current response to a local spin injection~[Fig.~\ref{fig:1} (b)]. 
For the two-dimensional Rashba model, which is an important example for the (I)SHE, we show that the charge current in such a configuration forms a vortex whose spatial profile indicates the strength of the SOC~\cite{lange2021}.

Specifically, we consider a Rashba system on a square lattice with one site coupled to a classical spin. The spin oscillates in the $xy$-plane and thereby locally induces a $z$-polarized spin current. 
We calculate the induced charge-current density analytically in the framework of response theory, obtaining a vortical structure for the dc component that 
can be regarded as the effect of spin precession and a spin-dependent force due to the Rashba SOC. 
The results are in agreement with a semiclassical wave-packet analysis~(WPA)~\cite{lange2021} as well as with numerical simulations.

As in the field of spintronics the lattice Rashba model has been less commonly used than the continuum one~\cite{bychkov1984}, we first derive the underlying tight-binding Hamiltonian in Sec.~\ref{sec:model}.
Considering the model on a square lattice, we define the charge and spin currents in Sec.~\ref{sec:charge_current} and Sec.~\ref{sec:spin_current}, respectively, by using the local $\mathrm{U}(1)$ and $\mathrm{SU}(2)$ gauge transformations. 
These definitions are consistent with the ordinary ones, in which the currents are introduced through the corresponding polarizations and their time derivatives.
In a next step, we construct the SOC as the generator of the spin current in Sec.~\ref{sec:soc}. 
Although this is different from the approach in Ref.~\onlinecite{ando1989}, it results in an equivalent Hamiltonian.
In Sec.~\ref{sec:vortex}, we finally consider the Rashba system with the classical spin and present the charge-current vortex generation by spin pumping.
Section~\ref{sec:summary} summarizes our work.
Appendix~\ref{apx:continuity} discusses the lattice versions of the continuity equations for charge and spin. 

\section{\label{sec:model}Tight-binding model}
We begin with the ordinary tight-binding model on the square lattice, which is given by

\begin{align}
\mathcal{H}_t
	& = - \frac{t}{2} \sum_{n,m} \Bigl(
			c^{\dagger}_{n,m} c^{}_{n+1,m}
			+ c^{\dagger}_{n,m} c^{}_{n-1,m}
\notag \\ & \hspace{4em}
			+ c^{\dagger}_{n,m} c^{}_{n,m+1}
			+ c^{\dagger}_{n,m} c^{}_{n,m-1}
			+ \mathrm{h.c.}
		\Bigr)
\label{eq:H_t}
,\end{align}
where $t > 0$ is the nearest-neighbor hopping parameter and $c_{n,m} = {}^{\mathrm{t}} (c_{n,m, \uparrow}, c_{n,m, \downarrow})$ is the spinor form of the electron operator for the site at $\bm{R}_{n,m} = n a \hat{x} + m a \hat{y}$  ($n$ and $m$ are integers, $a$ is the lattice constant). 
The factor $1/2$ is included to avoid double counting. 

By the Fourier transformation
\begin{align}
c^{}_{n,m}
	& = \frac{1}{\sqrt{N}} \sum_{\bk} c^{}_{\bk} e^{\zi \bk \cdot \bm{R}_{n,m}}
,
\label{eq:Fourier}%
\end{align}
where $N$ is the total site number, 
the Hamiltonian for periodic boundary conditions can be written as
\begin{align}
\mathcal{H}_t
	& = \sum_{\bk} T_{\bk} c^{\dagger}_{\bk} c^{}_{\bk}
\end{align}
with 
$T_{\bk} = - 2 t ( \cos k_x a + \cos k_y a )$. 
For $\bk \ll 2 \pi / a$, we can approximate the energy dispersion as $T_{\bk} = - 4 t + \hbar^2 k^2 / 2 \me$ with the Dirac constant $\hbar$, $k = |\bk | $ and $\frac{\hbar^2}{2 \me} \equiv t a^2$. 
In the long-wavelength limit, the system thus behaves like a free electron model with effective electron mass $\me$. 
Below, we introduce the charge- and spin-current operators for the lattice model and show explicitly that in the long-wavelength limit, they become equivalent to their counterparts for the free-electron model.

\section{\label{sec:charge_current}Charge-current operator}
The charge-current operator for the tight-binding model on the square lattice can be derived by exploiting the local $\mathrm{U}(1)$ gauge transformation
\begin{align}
c_{n,m} = U_{n,m} \bar{c}_{n,m}
, \qquad
U_{n,m} = e^{\zi \phi_{n,m}}
, \qquad
\phi_{n,m} \ll 1
\label{eq:U}
.\end{align}
Here, $\phi_{n,m} = \phi (\bm{R}_{n,m})$ and $\phi (\br)$ is a smooth real-valued function slowly-varying on the lattice constant scale.
When using the transformation~\eqref{eq:U} on the Hamiltonian~(\ref{eq:H_t}), combinations of unitary matrices for two different positions, such as $U_{n,m}^{\dagger} U^{}_{n+1,m}$, arise. 
Because $\phi_{n,m} \ll 1$, one can approximate 
\begin{align}
U_{n,m}^{\dagger} U_{n+1, m}
	& \simeq ( 1 - \zi \phi_{n,m}) ( 1 + \zi \phi_{n+1, m} )
\notag \\
	& \simeq 1 + \zi (\phi_{n+1, m} - \phi_{n,m}) .
\end{align}
Expanding
$\phi_{n+1,m} \simeq \phi_{n,m} + a \frac{\partial}{\partial x} \phi (\br) |_{\br = \bm{R}_{n,m}}$ then yields
\begin{align}
U_{n,m}^{\dagger} U^{}_{n+1, m}
	= 1 - \frac{\zi e a}{\hbar} A_x^{\mathrm{em}} (\bm{R}_{n,m})
,\end{align}
where $e$ is the elementary charge\footnote{To be exact, the coefficient $- \hbar / e$ in Eq.~(\ref{def:A_em}) is determined by the correspondence with the charge current operator in the continuum model.} and we have introduced the electromagnetic vector potential as
\begin{align}
A_x^{\mathrm{em}} (\br)
	& \equiv - \frac{\hbar}{e} \frac{\partial}{\partial x} \phi (\br) .
\label{def:A_em}
\end{align}
Similarly, we find
\begin{align}
U_{n,m}^{\dagger} U^{}_{n, m+1}
	= 1 - \frac{\zi e a}{\hbar} A_y^{\mathrm{em}} (\bm{R}_{n,m})
, 
\end{align}
where
\begin{align}
  A_y^{\mathrm{em}} (\br) & \equiv - \frac{\hbar}{e} \frac{\partial}{\partial y} \phi (\br) .
  \end{align}
The Hamiltonian can thus be written as $\mathcal{H}_t = \bar{\mathcal{H}}_t + \mathcal{H}_{\mathrm{em}}$, where $\bar{\mathcal{H}}_t$ is obtained by replacing $c^{(\dagger)}_{n,m}$ with $\bar{c}^{(\dagger)}_{n,m}$ in $\mathcal{H}_t$ 
and $\mathcal{H}_{\mathrm{em}}$ is composed of terms linear in the vector potential:
\begin{align}
\mathcal{H}_{\mathrm{em}}
	& = \frac{\zi e t a}{\hbar} \sum_{n,m} \Bigl(
			  \bar{c}^{\dagger}_{n,m} \frac{ \bar{c}^{}_{n+1,m} - \bar{c}^{}_{n-1,m} }{2} A_x^{\mathrm{em}} (\bm{R}_{n,m})
			- \mathrm{h.c.}
\notag \\ & \hspace{4em}
			+ \bar{c}^{\dagger}_{n,m} \frac{ \bar{c}^{}_{n,m+1} - \bar{c}^{}_{n,m-1} }{2} A_y^{\mathrm{em}} (\bm{R}_{n,m})
			- \mathrm{h.c.}
		\Bigr)
.\end{align}
We define the charge-current operator as the current coupled linearly with the vector potential, i.e., 
\begin{align}
j_x (\bm{R}_{n,m})
	& = \frac{\delta \mathcal{H}_t}{\delta A^{\mathrm{em}}_x (\bm{R}_{n,m})} \Bigg|_{\phi = 0}
\label{def1:jx}
, \\
j_y (\bm{R}_{n,m})
	& = \frac{\delta \mathcal{H}_t}{\delta A^{\mathrm{em}}_y (\bm{R}_{n,m})} \Bigg|_{\phi = 0}
\label{def1:jy}
,\end{align}
which leads to the following expressions:
\begin{align}
j_x (\bm{R}_{n,m})
	& = \frac{\zi e t a^2}{\hbar} \left(
		c^{\dagger}_{n,m} \frac{c^{}_{n+1,m} - c^{}_{n-1,m}}{2 a}
		- \mathrm{h. c.}
	\right)
\label{eq:jx}
, \\
j_y (\bm{R}_{n,m})
	& = \frac{\zi e t a^2}{\hbar} \left(
		c^{\dagger}_{n,m} \frac{c^{}_{n,m+1} - c^{}_{n,m-1}}{2 a}
		- \mathrm{h. c.}
	\right)
\label{eq:jy}
.\end{align}
In the continuum limit $c^{(\dagger)}_{n,m} \to c^{(\dagger)} (\br)$, we have
\begin{align}
\frac{c^{}_{n+1,m} - c^{}_{n-1,m}}{2 a}
	\to \frac{\partial c (\br)}{\partial x}
, \quad
\frac{c^{}_{n,m+1} - c^{}_{n,m-1}}{2 a}
	\to \frac{\partial c (\br)}{\partial y}
,\end{align}
which yields
\begin{align}
\bm{j} (\br)
	& = \frac{\zi e \hbar}{2 \me} \left( c^{\dagger} \bm{\nabla} c - (\bm{\nabla} c^{\dagger}) c \right) .
\label{eq:j-continuum}
\end{align}
Equation~(\ref{eq:j-continuum}) is consistent with the paramagnetic charge-current operator in the free electron model.

We note that the conventional definition of the total charge current is~\cite{mahan2000}
\begin{align}
\bm{J}
	& = \frac{\partial \bm{P}}{\partial t}
	= - \frac{\zi}{\hbar} [\bm{P}, \mathcal{H}_t]
\label{def:J}
,\end{align}
where $\bm{P} = - e \sum_{n,m} \bm{R}_{n,m} c^{\dagger}_{n,m} c^{}_{n,m}$ is the (total) polarization. 
By considering that the total charge current is the sum of the local charge currents, 
$\bm{J} = \sum_{n,m} \bm{j} (\bm{R}_{n,m})$, 
expressions equivalent to Eqs.~(\ref{eq:jx}) and (\ref{eq:jy}) are obtained.

\section{\label{sec:spin_current}Spin-current operator}
In this section, we derive the spin-current operator for the square lattice, by using the local $\mathrm{SU}(2)$ gauge transformation
\begin{align}
c_{n,m} = V_{n,m} \tilde{c}_{n,m}
, \qquad
V_{n,m} = e^{\zi \theta \bm{\omega}_{n,m} \cdot \bm{\sigma}}
, \qquad
\theta \ll 1
\label{eq:V}
,\end{align}
where $\bm{\sigma} = (\sigma^x, \sigma^y, \sigma^z)$ is the vector of Pauli matrices and $|\omega_{n,m}| = 1$. Furthermore, $\bm{\omega}_{n,m} = \bm{\omega} (\bm{R}_{n,m})$, with 
a smooth real-valued vector field $\bm{\omega} (\br) = (\omega^x (\br), \omega^y (\br), \omega^z (\br))$ slowly-varying on the lattice-constant scale. Equation~\eqref{eq:V} describes a spin rotation around the vector $\omega_{n,m}$ with the angle $\theta$. We can approximate $V_{n,m} \simeq 1 + \zi \theta \bm{\omega}_{n,m} \cdot \bm{\sigma}$, since $\theta \ll 1$. 
The gauge transformation of the Hamiltonian~(\ref{eq:H_t}) then follows in a similar way to the $\mathrm{U}(1)$ gauge transformation in Sec.~\ref{sec:charge_current}. 
To first order in $\theta$:
\begin{align}
V_{n,m}^{\dagger} V^{}_{n+1,m}
	& = 1 + \zi a \bm{A}_x (\bm{R}_{n,m}) \cdot \frac{\bm{\sigma}}{2}
, \\
V_{n,m}^{\dagger} V^{}_{n,m+1}
	& = 1 + \zi a \bm{A}_y (\bm{R}_{n,m}) \cdot \frac{\bm{\sigma}}{2}
,\end{align}
where we have introduced the $\mathrm{SU}(2)$ gauge potential
\begin{align}
\bm{A}_x (\br)
	& = 2 \theta \bm{\omega} (\br) \times \frac{\partial}{\partial x} \bm{\omega} (\br)
, \\
\bm{A}_y (\br)
	& = 2 \theta \bm{\omega} (\br) \times \frac{\partial}{\partial y} \bm{\omega} (\br)
.\end{align}
The Hamiltonian can now be rewritten as $\mathcal{H}_t = \tilde{\mathcal{H}}_t + \mathcal{H}_{A}$, where $\tilde{\mathcal{H}}_t$ is the Hamiltonian obtained by replacing $c^{(\dagger)}_{n,m}$ with $\tilde{c}^{(\dagger)}_{n,m}$ in $\mathcal{H}_t$, and 
\begin{align}
\mathcal{H}_A
	& = - \frac{\zi t a}{4} \sum_{n,m} \Bigl(
			  \tilde{c}^{\dagger}_{n,m} \bm{\sigma} \frac{ \tilde{c}^{}_{n+1,m} - \tilde{c}^{}_{n-1,m} }{2} \cdot \bm{A}_x (\bm{R}_{n,m})
			- \mathrm{h.c.}
\notag \\ & \hspace{4em}
			+ \tilde{c}^{\dagger}_{n,m} \bm{\sigma} \frac{ \tilde{c}^{}_{n,m+1} - \bm{\sigma} \tilde{c}^{}_{n,m-1} }{2} \cdot \bm{A}_y (\bm{R}_{n,m})
			- \mathrm{h.c.}
		\Bigr)
\label{eq:H_A}
.\end{align}
We define the spin-current operator as the current coupled linearly with the $\mathrm{SU}(2)$ gauge potential, i.e.,
\begin{align}
j_{\mathrm{s}, x}^{\alpha} (\bm{R}_{n,m})
	& = \frac{\delta \mathcal{H}_t}{\delta A^{\alpha}_x (\bm{R}_{n,m})} \Bigg|_{\theta = 0}
\label{def1:jsx}
, \\
j_{\mathrm{s}, y}^{\alpha} (\bm{R}_{n,m})
	& = \frac{\delta \mathcal{H}_t}{\delta A^{\alpha}_y (\bm{R}_{n,m})} \Bigg|_{\theta = 0}
\label{def1:jsy}
,\end{align}
with $\alpha \in \{ x, y, z \}$ being the spin index, which yields
\begin{align}
j_{\mathrm{s}, x}^{\alpha} (\bm{R}_{n,m})
	& = - \frac{\zi t a^2}{2} \left(
		c^{\dagger}_{n,m} \sigma^{\alpha} \frac{ c^{}_{n+1,m} - c^{}_{n-1,m} }{2 a}
		- \mathrm{h.c.}
	\right)
\label{eq:js_x}
, \\
j_{\mathrm{s}, y}^{\alpha} (\bm{R}_{n,m})
	& = - \frac{\zi t a^2}{2} \left(
		c^{\dagger}_{n,m} \sigma^{\alpha} \frac{ c^{}_{n,m+1} - c^{}_{n,m-1} }{2 a}
		- \mathrm{h.c.}
	\right)
\label{eq:js_y}
.\end{align}

In the continuum limit $c^{(\dagger)}_{n,m} \to c^{(\dagger)} (\br)$, the spin-current operator becomes
\begin{align}
j_{\mathrm{s}, x}^{\alpha} (\br)
	& = \frac{\hbar}{2} \frac{\hbar}{2 \me \zi} \left(
		c^{\dagger} \sigma^{\alpha} \frac{\partial c}{\partial x}
		- \frac{\partial c^{\dagger}}{\partial x} \sigma^{\alpha} c
	\right)
, \\
j_{\mathrm{s}, y}^{\alpha} (\br)
	& = \frac{\hbar}{2} \frac{\hbar}{2 \me \zi} \left(
		c^{\dagger} \sigma^{\alpha} \frac{\partial c}{\partial y}
		- \frac{\partial c^{\dagger}}{\partial y} \sigma^{\alpha} c
	\right)
,\end{align}
which is again consistent with the  expression for the free electron model.

Let us briefly comment on the definition of the spin-current operator proposed by Shi \textit{et al.}~\cite{shi2006}. 
They introduce the spin-current operator as the time derivative of the spin-displacement operator $\br s^z$:
\begin{align}
\bm{j}_{\mathrm{s}}^z (\br)
	& = \frac{\mathrm{d} (\br s^z)}{\mathrm{d} t}.
\end{align}
This definition can be regarded as an 
extension of Eq.~(\ref{def:J}). 
To see this, we define the total spin polarization as
$\bm{P}^z  = \frac{\hbar}{2} \sum_{n,m} \bm{R}_{n,m} c^{\dagger}_{n,m} \sigma^z c^{}_{n,m}$
and the total spin current as its time derivative:
\begin{align}
\bm{J}_{\mathrm{s}}^z
	& = \frac{\partial \bm{P}^z}{\partial t}
	= - \frac{\zi}{\hbar} [ \bm{P}^z, \mathcal{H}_t ]
\label{def:Jz_shi}
.\end{align}
Comparing with 
$\bm{J}_{\mathrm{s}}^z = \sum_{n,m} \bm{j}_{\mathrm{s}}^z (\bm{R}_{n,m})$
then leads to Eqs.~(\ref{eq:js_x}) and (\ref{eq:js_y}) for $\alpha = z$.

\section{\label{sec:soc}Spin-orbit coupling}
Although the Rashba SOC in the square lattice has been derived from the atomic SOC with inversion asymmetry~\cite{ando1989}, we here introduce it 
in a different way and obtain the same result. 
In the continuum model, the SOC is the generator of the intrinsic spin current, i.e., $\mathcal{H}_{\mathrm{soc}} = ( 2 \me / \hbar^2 )\sum_{i, \alpha} \lambda_{i}^{\alpha} \int \mathrm{d}\br j_{\mathrm{s}, i}^{\alpha} (\br)$ with the coefficient $\lambda_i^{\alpha}$ specifying the SOC amplitude~\cite{kikuchi2016}. 
We extend this definition of SOC to the lattice model as
\begin{align}
\mathcal{H}_{\mathrm{soc}}
	& = \sum_{i, \alpha} \frac{\lambda_i^{\alpha}}{t a^2} \sum_{n,m} j_{\mathrm{s}, i}^{\alpha} (\bm{R}_{n,m})
\label{eq:soc}
,\end{align}
where $j_{\mathrm{s}, i}^{\alpha} (\bm{R}_{n,m})$ is the spin-current operator with direction $i \in \{x,y\}$ and spin index $\alpha \in\{ x, y, z \}$. 
The SOC amplitude $\lambda_i^{\alpha}$ in the case of the Rashba SOC is 
\begin{align}
\lambda_{\mathrm{R}, i}^{\alpha}
	& = \alpha_{\R} \epsilon^{i \alpha z}
	= \alpha_{\R}
	\begin{pmatrix}
		0
	&	1
	&	0
	\\	- 1
	&	0
	&	0
	\\	0
	&	0
	&	0
	\end{pmatrix}_{i \alpha}
\label{eq:lambda_R}
,\end{align}
where $\epsilon^{i j k}$ is the antisymmetric tensor and $\alpha_{\R}$ is the Rashba parameter.
Substituting Eq.~(\ref{eq:lambda_R}) into Eq.~(\ref{eq:soc}) results in
\begin{align}
\mathcal{H}_{\mathrm{R}}
	& = - \frac{\zi \alpha_{\R}}{2} \sum_{n,m} \Bigl(
		c^{\dagger}_{n,m} \sigma^{y} \frac{ c^{}_{n+1,m} - c^{}_{n-1,m} }{2 a}
		- \mathrm{h.c.}
\notag \\[-1ex] & \hspace{4.5em}
		- c^{\dagger}_{n,m} \sigma^{x} \frac{ c^{}_{n,m+1} - c^{}_{n,m-1} }{2 a}
		+ \mathrm{h.c.}
	\Bigr)
\label{eq:Rashba}
\end{align}
or, in the momentum representation, 
\begin{align}
\mathcal{H}_{\R}
	& = \frac{\alpha_{\R}}{a} \sum_{\bk} c^{\dagger}_{\bk} \left( \sigma^y \sin k_x a - \sigma^x \sin k_y a \right) c^{}_{\bk} \nonumber \\
&= \sum_{\bk} c^{\dagger}_{\bk} (\bm{\lambda}_{\bk} \cdot \bm{\sigma}) c^{}_{\bk},
\end{align}
with $\bm{\lambda}_{\bk} = (\alpha_{\R}/a) (- \sin k_y a, \sin k_x a, 0)$. 
Considering the long-wavelength limit $k \ll 2 \pi / a$, $(\sin k_x a)/a \simeq k_x$ and $(\sin k_y a)/a \simeq k_y$, we obtain
\begin{align}
\mathcal{H}_{\mathrm{Rashba}}
&= \alpha_{\R} \sum_{\bk} c^{\dagger}_{\bk} (\bk \times \bm{\sigma})^z c^{}_{\bk},
\end{align}
which is the well-known Rashba Hamiltonian in the continuum.

While we focus here on the Rashba model, the above procedure can be used for other types of SOC too. 
From the knowledge of the continuum model, the Dresselhaus SOC amplitude may be shown to be
\begin{align}
\lambda_{\mathrm{D}, i}^{\alpha}
	& = \beta_{\mathrm{D}}
	\begin{pmatrix}
		1
	&	0
	&	0
	\\	0
	&	-1
	&	0
	\\	0
	&	0
	&	0
	\end{pmatrix}_{i \alpha}
,\end{align}
where $\beta_{\mathrm{D}}$ is the Dresselhaus SOC strength. In the case of the Weyl SOC for the square lattice, we have 
\begin{align}
\lambda_{\mathrm{W}, i}^{\alpha}
	& = \gamma_{\mathrm{W}}
	\begin{pmatrix}
		1
	&	0
	&	0
	\\	0
	&	1
	&	0
	\\	0
	&	0
	&	0
	\end{pmatrix}_{i \alpha}
,\end{align}
where $\gamma_{\mathrm{W}}$ is the strength of the Weyl SOC.

Next, we derive the charge- and spin-current operators in the presence of Rashba SOC. 
By applying the local $\mathrm{U}(1)$ gauge transformation~(\ref{eq:U}) to the Rashba Hamiltonian~(\ref{eq:Rashba}), we get $\mathcal{H}_{\R} = \bar{\mathcal{H}}_{\R} + \mathcal{H}'_{\mathrm{em}}$, where $\bar{\mathcal{H}}_{\R}$ is the Hamiltonian obtained by replacing $c^{(\dagger)}_{n,m}$ with $\bar{c}^{(\dagger)}_{n,m}$ in $\mathcal{H}_{\R}$, and
\begin{align}
\mathcal{H}'_{\mathrm{em}}
	& = - \frac{e \alpha_{\R}}{4 \hbar} \sum_{n,m} \Bigl(
		  \bar{c}^{\dagger}_{n,m} \sigma^{y} \bar{c}^{}_{n+1,m} A^{\mathrm{em}}_x (\bm{R}_{n,m})
		+ \mathrm{h.c.}
\notag \\[-1ex] & \hspace{5em}
		+ \bar{c}^{\dagger}_{n,m} \sigma^{y} \bar{c}^{}_{n-1,m} A^{\mathrm{em}}_x (\bm{R}_{n,m})
		+ \mathrm{h.c.}
\notag \\ & \hspace{5em}
		- \bar{c}^{\dagger}_{n,m} \sigma^{x} \bar{c}^{}_{n,m+1} A^{\mathrm{em}}_y (\bm{R}_{n,m})
		- \mathrm{h.c.}
\notag \\ & \hspace{5em}
		- \bar{c}^{\dagger}_{n,m} \sigma^{x} \bar{c}^{}_{n,m-1} A^{\mathrm{em}}_y (\bm{R}_{n,m})
		- \mathrm{h.c.}
	\Bigr)
.\end{align}
By defining the charge-current operator analogously to Eqs.~\eqref{def1:jx} and \eqref{def1:jy}, 
we get the following expressions:
\begin{align}
j'_{x} (\bm{R}_{n,m})
	& = - \frac{e \alpha_{\R}}{2 \hbar} \left(
		c^{\dagger}_{n,m} \sigma^{y} \frac{ c^{}_{n+1,m} + c^{}_{n-1,m} }{2}
		+ \mathrm{h.c.}
	\right)
\label{eq:jx_R}
, \\
j'_{y} (\bm{R}_{n,m})
	& = + \frac{e \alpha_{\R}}{2 \hbar} \left(
		c^{\dagger}_{n,m} \sigma^{x} \frac{ c^{}_{n,m+1} + c^{}_{n,m-1} }{2}
		+ \mathrm{h.c.}
	\right)
\label{eq:jy_R}
.\end{align}
In the continuum limit, where $(c^{}_{n+1,m} + c^{}_{n-1,m}) / 2 \to c (\br)$, we confirm the correspondence to the anomalous charge-current operators in the free electron model with Rashba SOC,
\begin{align}
\bm{j}' (\br)
	& = \frac{e \alpha_{\R}}{\hbar} c^{\dagger} (\hat{z} \times \bm{\sigma}) c
\label{eq:j-continuum_R}
,\end{align}
where $c^{(\dagger)} = c^{(\dagger)} (\br)$.
Note that the anomalous charge-current operator~(\ref{eq:j-continuum_R}) is proportional to the spin-density operator $\bm{s} (\br) = c^{\dagger} \bm{\sigma} c$ as $\bm{j}' (\br) = (e \alpha_{\R} / \hbar) \hat{z} \times \bm{s} (\br)$.

The spin current is derived by using the local $\mathrm{SU}(2)$ gauge transformation~(\ref{eq:V}) on the Rashba Hamiltonian~(\ref{eq:Rashba}). 
We get $\mathcal{H}_{\R} = \tilde{\mathcal{H}}_{\R} + \mathcal{H}'_{A}$, where $\tilde{\mathcal{H}}_{\R}$ is the Hamiltonian obtained by replacing $c^{(\dagger)}_{n,m}$ with $\tilde{c}^{(\dagger)}_{n,m}$ and $\sigma^{\alpha}$ ($\alpha = x, y$) with $\tilde{\sigma}^{\alpha} \equiv V_{n,m}^{\dagger} \sigma^{\alpha} V_{n,m}$ in $\mathcal{H}_{\R}$, and
\begin{align}
\mathcal{H}'_A
	& = \frac{\alpha_{\R}}{2} \sum_{n,m} \Bigl(
		  \tilde{c}^{\dagger}_{n,m} \tilde{\sigma}^{y} \sigma^{\alpha} \tilde{c}^{}_{n+1,m} A^{\alpha}_x (\bm{R}_{n,m})
		+ \mathrm{h.c.}
\notag \\[-1ex] & \hspace{3.5em}
		+ \tilde{c}^{\dagger}_{n,m} \tilde{\sigma}^{y} \sigma^{\alpha} \tilde{c}^{}_{n-1,m} A^{\alpha}_x (\bm{R}_{n,m})
		+ \mathrm{h.c.}
\notag \\ & \hspace{3.5em}
		- \tilde{c}^{\dagger}_{n,m} \tilde{\sigma}^{x} \sigma^{\alpha} \tilde{c}^{}_{n,m+1} A^{\alpha}_y (\bm{R}_{n,m})
		- \mathrm{h.c.}
\notag \\ & \hspace{3.5em}
		- \tilde{c}^{\dagger}_{n,m} \tilde{\sigma}^{x} \sigma^{\alpha} \tilde{c}^{}_{n,m-1} A^{\alpha}_y (\bm{R}_{n,m})
		- \mathrm{h.c.}
	\Bigr)
.\end{align}
The spin-current operator for the Rashba Hamiltonian is then defined as
\begin{align}
j_{\mathrm{s}, x}^{\prime \alpha} (\bm{R}_{n,m})
	& = \frac{1}{2} \frac{\delta \mathcal{H}_{\R}}{\delta A^{\alpha}_x (\bm{R}_{n,m})} \Bigg|_{\theta = 0}
\label{def1:jsx_R}
, \\
j_{\mathrm{s}, y}^{\prime \alpha} (\bm{R}_{n,m})
	& = \frac{1}{2} \frac{\delta \mathcal{H}_{\R}}{\delta A^{\alpha}_y (\bm{R}_{n,m})} \Bigg|_{\theta = 0}
\label{def1:jsy_R}
,\end{align}
so that we have
\begin{align}
j_{\mathrm{s}, x}^{\prime \alpha} (\bm{R}_{n,m})
	& = j_{\mathrm{s}, x}^{(1), \alpha} (\bm{R}_{n,m}) + j_{\mathrm{s}, x}^{(2), \beta} (\bm{R}_{n,m}) \epsilon^{y \alpha \beta}
\label{eq:jsx_R}
, \\
j_{\mathrm{s}, y}^{\prime \alpha} (\bm{R}_{n,m})
	& = j_{\mathrm{s}, y}^{(1), \alpha} (\bm{R}_{n,m}) + j_{\mathrm{s}, y}^{(2), \beta} (\bm{R}_{n,m}) \epsilon^{x \alpha \beta}
\label{eq:jsy_R}
,\end{align}
where
\begin{align}
j_{\mathrm{s}, x}^{(1), y} (\bm{R}_{n,m})
	& = \frac{\alpha_{\R}}{4} \left(
		c^{\dagger}_{n,m} \frac{ c^{}_{n+1,m} + c^{}_{n-1,m} }{2}
		+ \mathrm{h.c.}
	\right)
, \\
j_{\mathrm{s}, y}^{(1), x} (\bm{R}_{n,m})
	& = - \frac{\alpha_{\R}}{4} \left(
		c^{\dagger}_{n,m} \frac{ c^{}_{n,m+1} + c^{}_{n,m-1} }{2}
		+ \mathrm{h.c.}
	\right)
\end{align}
and the other components of $j_{\mathrm{s}, i}^{(1), \alpha}$ are zero, and
\begin{align}
j_{\mathrm{s}, x}^{(2), \beta} (\bm{R}_{n,m})
	& = \frac{\zi \alpha_{\R}}{4} \left(
		c^{\dagger}_{n,m} \sigma^{\beta} \frac{ c^{}_{n+1,m} + c^{}_{n-1,m} }{2}
		- \mathrm{h.c.}
	\right)
, \\
j_{\mathrm{s}, y}^{(2), \beta} (\bm{R}_{n,m})
	& = - \frac{\zi \alpha_{\R}}{4} \left(
		c^{\dagger}_{n,m} \sigma^{\beta} \frac{ c^{}_{n,m+1} + c^{}_{n,m-1} }{2}
		- \mathrm{h.c.}
	\right)
.\end{align}
In the continuum limit, we obtain
\begin{align}
j_{\mathrm{s}, x}^{(1), y} (\br)
	= - j_{\mathrm{s}, y}^{(1), x} (\br)
	= \frac{1}{2} \alpha_{\R} c^{\dagger} c
,\end{align}
or $j_{\mathrm{s}, i}^{(1), \alpha} (\br) = \epsilon^{i \alpha z} (\alpha_{\R}/2) c^{\dagger} c$, and $j_{\mathrm{s}, i}^{(2), \beta} (\br) = 0$, which means
\begin{align}
\bm{j}_{\mathrm{s}}^{\prime \alpha} (\br)
	& = \bm{j}_{\mathrm{s}}^{(1), \alpha} (\br)
	= \frac{\alpha_{\R}}{2} (\hat{\alpha} \times \hat{z}) c^{\dagger} c
\label{eq:js_R_continuum}
.\end{align}
This anomalous spin-current operator is proportional to the particle-density operator $n = c^{\dagger} c$.

Note that in our definition of the spin-current operator for the Rashba Hamiltonian [see Eqs.~(\ref{def1:jsx_R}) and (\ref{def1:jsy_R})] an additional factor $1/2$ is included. 
The factor originates from the fact that the Rashba SOC is the generator of the intrinsic spin current.
Since the intrinsic spin current is also generated by the $\mathrm{SU}(2)$ gauge potential as shown in Eq.~(\ref{eq:H_A}), the SOC amplitude $\lambda_i^{\alpha}$ acts as an $\mathrm{SU}(2)$ gauge potential in the continuum model:
\begin{align}
\mathcal{H}'
	& = \sum_{i,\alpha} \int \mathrm{d}\br \tilde{j}_{\mathrm{s},i}^{\alpha} (\br) ( \bm{A}_i - \mathcal{R}^{-1} \bm{\lambda}_i )^{\alpha}
\notag \\ &
	+ \frac{1}{2} \sum_{i,\alpha} \int \mathrm{d}\br \frac{1}{2} \tilde{n} (\br) ( \bm{A}_i - \mathcal{R}^{-1} \bm{\lambda}_i )^{\alpha} ( \bm{A}_i - \mathcal{R}^{-1} \bm{\lambda}_i )^{\alpha}
\notag \\ & 
	+ \mathcal{O} ( \{\lambda^{\alpha}_i\}^2 )
.\end{align}
Here, $\tilde{j}_{\mathrm{s},i}^{\alpha} (\br)$ and $\tilde{n} (\br)$ are the spin-current and electron-density operators described by the field operator $\tilde{c}^{(\dagger)}$. Moreover, we have introduced the vector forms of the $\mathrm{SU}(2)$ gauge potential $\bm{A}_i = (A_i^x, A_i^y, A_i^z)$ and the SOC amplitude $\bm{\lambda}_i = (\lambda_i^x, \lambda_i^y, \lambda_i^z)$ with the rotational matrix $\mathcal{R}$ defined by $U^{\dagger} \bm{\sigma} U = \mathcal{R} \bm{\sigma}$.
We need to count the order of such a gauge potential including the SOC amplitude, which means that the terms proportional to both of the gauge potential and of the SOC amplitude correspond to the second order. 
The factor $1/2$ difference in the definitions will be clarified in Appendix~\ref{apx:continuity} discussing the continuity equations.

We now comment on the connection between the definition in Ref. \onlinecite{shi2006} and Eqs. (\ref{eq:jsx_R}) and (\ref{eq:jsy_R}).
Calculating Eq.~(\ref{def:Jz_shi}) for the Rashba Hamiltonian~(\ref{eq:Rashba}), we find a torque term $T^{\alpha}$ arises in addition to the expressions~(\ref{eq:jsx_R}) and (\ref{eq:jsy_R}),
\begin{align}
T^{\alpha}
	& = \frac{\alpha_{\R}}{t a^2} \sum_{n, m} \sum_{i,j,\beta} \bm{R}_{n, m} \epsilon^{i j z} \epsilon^{\alpha j \beta} j_{\mathrm{s}, i}^{\beta} (\bm{R}_{n,m})
.\end{align}
Especially, $\alpha = z$,
\begin{align}
T^z
	& = - \frac{\alpha_{\R}}{t a^2} \sum_{n, m} \bm{R}_{n, m} \left(
		j_{\mathrm{s}, x}^{x} (\bm{R}_{n,m})
		+ j_{\mathrm{s}, y}^{y} (\bm{R}_{n,m})
	\right)
.\end{align}
We emphasize that we have defined the spin-current operator that corresponds to the spin operator $s^{\alpha}$ not to the spin-displacement operator $\br s^{\alpha}$, so that, as shown in Appendix~\ref{apx:continuity}, the continuity equation for spin is not fulfilled.  Instead, a continuity equation holds for the total spin-displacement operator. 
These definitions are equivalent in the absence of SOCs, but they generally lead to different expressions in spin-orbit coupled systems.

For a meaningful discussion of spin transport, it is important that the spin current is defined properly.
One possible proper way to define the spin current operator is based on the local $\mathrm{SU}(2)$ gauge transformation, and another is based on the continuity equation for spin, both of which should be equivalent.
Sometimes, a spin current operator is naively introduced~\cite{rashba2003} as $\bar{j}_{\mathrm{s}, i}^{\alpha} (\br) = ( s^{\alpha} v_i + v_i s^{\alpha} )/2$, where $v_i$ is the velocity operator.
However, the operator $\bar{j}_{\mathrm{s}, i}^{\alpha} (\br)$ in the presence of the Rashba SOC is not same as our expression~(\ref{eq:js_R_continuum}), where the factor $1/2$ is different.
Moreover, when one uses the continuity equation for spin to discuss spin accumulation in finite systems, the boundary conditions need to be taken into account~\cite{shi2006,nomura2005,tse2005}.

\section{\label{sec:vortex}Charge-current-vortex generation by spin pumping}
\begin{figure}[b]
\centering
\includegraphics[width=\linewidth,clip]{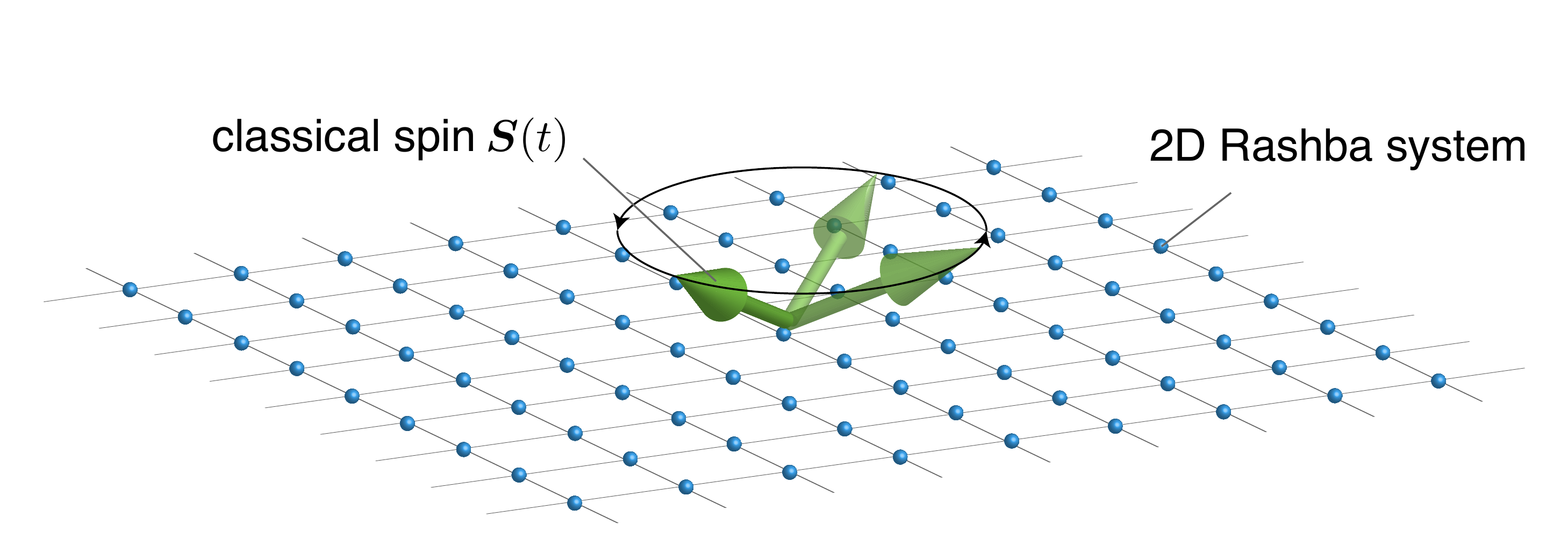}
\caption{\label{fig:spin-pump}Schematic figure of the local spin pumping in a Rashba model on a square lattice. The time-dependent classical spin $\bm{S} (t)$ at the origin is coupled to the electron spin through the exchange interaction. The dynamics of the classical spin induces a charge-current vortex.}
\end{figure}
After deriving the SOC Hamiltonian, we are now able to discuss the charge-current-vortex generation by spin injection via spin pumping. 
We consider a tight-binding model on the square lattice with Rashba SOC, which is coupled to a classical spin at the origin $\bm{R}_{0,0}$ through the exchange coupling
\begin{align}
\mathcal{H}_{S} (t)
	& = - J_s \bm{S} (t) \cdot \left( c^{\dagger}_{0,0} \bm{\sigma} c_{0,0}^{} \right)
.\end{align} 
We assume that the classical spin has the magnitude $S$ and precesses with the angular frequency $\omega$ in the $xy$-plane:
\begin{align}
\bm{S} (t)
	& = S
		\begin{pmatrix}
			\cos \omega t
		\\	\sin \omega t
		\\	0
		\end{pmatrix}
, \qquad
\dot{\bm{S}} (t)
	= S \omega
	\begin{pmatrix}
		- \sin \omega t
	\\	\cos \omega t
	\\	0
	\end{pmatrix}
\label{eq:S_dynamics}
.\end{align}
While we can do the following calculation also for $S_z \neq 0$, the dc component of the charge current density does not change qualitatively from the $S_z = 0$ case, so that we set $S_z = 0$ for simplicity. 

The classical spin dynamics generates a charge current through the Rashba SOC.
We now evaluate the charge-current density within response theory (see Appendix A in Ref.~\onlinecite{fujimoto2019} for details):
\begin{align}
J_i (\br, \omega)
	& = \chi_i^{\alpha} (\br, \omega) S (\omega)
\notag \\ &
	+ \int_{-\infty}^{\infty} \frac{\mathrm{d}\omega'}{2 \pi} \chi_{i}^{\alpha \beta} (\br, \omega, \omega') S^{\alpha} (\omega - \omega') S^{\beta} (\omega')
\notag \\ &
	+ \cdots
\end{align}
In the following, we focus on the dc component of the response in the lowest order with respect to $\bm{S} (\omega)$, which is 
\begin{align}
\vartheta_i^{\alpha \beta} (\br)
	& = \lim_{\omega' \to 0}
		\frac{\chi_i^{\alpha \beta} (\br, 0, \omega') - \chi_i^{\alpha \beta} (\br, 0, 0)}{- \zi \omega'}
.\end{align} 
The response coefficient $\chi_{i}^{\alpha \beta} (\br, \omega, \omega')$ is obtained from the correlation function in the Matsubara representation,
\begin{align}
X_i^{\alpha \beta} (\br, \zi \omega_{\lambda}, \zi \omega_{\lambda'})
	& = \frac{J_s^2}{2 N} \sum_{\bq} e^{\zi \bq \cdot \br} \iint_0^{\beta} \mathrm{d}\tau \mathrm{d}\tau' e^{\zi \omega_{\lambda} (\tau-\tau') + \zi \omega_{\lambda'} \tau'}
\notag \\ & \hspace{-1em} \times
		\left\langle \mathrm{T}_{\tau, \tau'} \{ J_i (\bq, \tau) s^{\alpha} (0, \tau') s^{\beta} (0, 0) \} \right\rangle
\end{align}
by taking the analytic continuations $\zi \omega_{\lambda} \to \hbar \omega + 2 \zi 0$ and $\zi \omega_{\lambda'} \to \hbar \omega' + \zi 0$:
\begin{align}
X_i^{\alpha \beta} (\br, \hbar \omega + 2 \zi 0, \hbar \omega' + \zi 0)
	& = \chi_{i}^{\alpha \beta} (\br, \omega, \omega')
.\end{align}
Here, $J_i (\bq, \tau)$ and $s^{\alpha} (0, \tau)$ are the charge-current operator and the spin operator in the Heisenberg picture, $\mathrm{T}_{\tau, \tau'}$ is the time-ordering operator, $\beta = 1 / k_{\mathrm{B}} T$ is the inverse temperature, and $\langle \cdots \rangle$ is the thermal average.
The corresponding operators in the Schr\"{o}dinger picture are 
\begin{align}
J_i (\bq)
	& = \sum_{\bk} c^{\dagger}_{\bk-\frac{\bq}{2}} ( v_{\bk, i} + v'_{\bk, i} ) c^{}_{\bk+\frac{\bq}{2}},
\end{align}
with the normal and anomalous velocities
$v_{\bk, i} = - \frac{2 e a t}{\hbar} \sin k_i a$
and 
$v'_{\bk, i} = - \frac{e \alpha_{\R}}{\hbar} \epsilon^{z i \beta} \sigma^{\beta} \cos k_i a$,
and 
\begin{align}
s^{\alpha} (0)
	& = c^{\dagger}_{0,0} \sigma^{\alpha} c^{}_{0,0}
	= \frac{1}{N} \sum_{\bk, \bk'} c^{\dagger}_{\bk'} \sigma^{\alpha} c^{}_{\bk}
.\end{align}
After some straightforward calculations, we find 
\begin{align}
\vartheta_i^{\alpha \beta} (\br)
	& = \frac{J_s^2}{N^3} \sum_{\bq} e^{\zi \bq \cdot \br} \sum_{\bk, \bk'}
		\frac{\zi \hbar}{2 \pi} \int_{-\infty}^{\infty} \mathrm{d}\epsilon \left( - \frac{\partial f}{\partial \epsilon} \right)
\notag \\ & \hspace{-3mm} \times
		\mathrm{tr} \left[
			\left( v_{\bk, i} + v'_{\bk, i} \right) G^{\R}_{\bk+\frac{\bq}{2}} (\epsilon)
			\sigma^{\alpha} \Im [ G^{\R}_{\bk'} (\epsilon) ]
			\sigma^{\beta} G^{\A}_{\bk-\frac{\bq}{2}} (\epsilon)
\right. \notag \\ & \hspace{-2mm} \left.
			- \left( v_{\bk, i} + v'_{\bk, i} \right) G^{\R}_{\bk+\frac{\bq}{2}} (\epsilon)
			\sigma^{\beta} \Im [ G^{\R}_{\bk'} (\epsilon) ]
			\sigma^{\alpha} G^{\A}_{\bk-\frac{\bq}{2}} (\epsilon)
		\right]
\notag \\ & \hspace{-3mm}
	- \frac{J_s^2}{N^3} \sum_{\bq} e^{\zi \bq \cdot \br} \sum_{\bk, \bk'} \frac{\hbar}{4 \pi} \int_{-\infty}^{\infty} \mathrm{d}\epsilon f (\epsilon)
\notag \\ & \hspace{-3mm} \times
		\mathrm{tr} \left[
			\left( v_{\bk, i} + v'_{\bk, i} \right) G^{\R}_{\bk+\frac{\bq}{2}} (\epsilon)
			\sigma^{\alpha} \bigl( \partial_{\epsilon} G^{\R}_{\bk'} (\epsilon) \bigr)
			\sigma^{\beta} G^{\R}_{\bk-\frac{\bq}{2}} (\epsilon)
\right. \notag \\ & \hspace{-2mm} \left.
			- \left( v_{\bk, i} + v'_{\bk, i} \right) G^{\R}_{\bk+\frac{\bq}{2}} (\epsilon)
			\sigma^{\beta} \bigl( \partial_{\epsilon} G^{\R}_{\bk'} (\epsilon) \bigr)
			\sigma^{\alpha} G^{\R}_{\bk-\frac{\bq}{2}} (\epsilon)
\right. \notag \\ & \hspace{-2mm} \left.
			- \left( v_{\bk, i} + v'_{\bk, i} \right) G^{\A}_{\bk+\frac{\bq}{2}} (\epsilon)
			\sigma^{\alpha} \bigl( \partial_{\epsilon} G^{\A}_{\bk'} (\epsilon) \bigr)
			\sigma^{\beta} G^{\A}_{\bk-\frac{\bq}{2}} (\epsilon)
\right. \notag \\ & \hspace{-2mm} \left.
			+ \left( v_{\bk, i} + v'_{\bk, i} \right) G^{\A}_{\bk+\frac{\bq}{2}} (\epsilon)
			\sigma^{\beta} \bigl( \partial_{\epsilon} G^{\A}_{\bk'} (\epsilon) \bigr)
			\sigma^{\alpha} G^{\A}_{\bk-\frac{\bq}{2}} (\epsilon)
		        \right]
                \label{eq:long}
,\end{align}
where $f (\epsilon)$ is the Fermi-Dirac distribution function, and $G^{\R/\A}_{\bk}$ is the retarded/advanced Green function defined by  $G^{\R/\A}_{\bk}(\epsilon) = G_{\bk} (\epsilon \pm \zi \gamma)$ and 
\begin{align}
G_{\bk} (z)
	& = \frac{ z + \mu - T_{\bk} + \bm{\lambda}_{\bk} \cdot \bm{\sigma} }{D_{\bk} (z)}
, \\
D_{\bk} (z)
	& = (z + \mu - T_{\bk})^2 - |\bm{\lambda}_{\bk}|^2
.\end{align}
Here, $\mu$ is the chemical potential, and $\gamma$ is the level broadening.
The terms proportional to $f (\epsilon)$ are expected to be negligible, since $\Re [1/D_{\bk} (\epsilon+\zi\gamma)] \approx 0$.
We therefore may simplify Eq.~\eqref{eq:long} to
\begin{align}
\vartheta_i^{\alpha \beta} (\br)
	& = - \frac{\hbar J_s^2}{4 N^2} \epsilon^{\alpha \beta z}
		\int_{-\infty}^{\infty} \mathrm{d}\epsilon \left( - \frac{\partial f}{\partial \epsilon} \right) \nu (\epsilon)
	\sum_{\bp, \bq} e^{\zi (\bp - \bq) \cdot \br}
\notag \\ & \hspace{1em} \times
	\mathrm{tr} \left[
			\left( v_{\frac{\bp+\bq}{2}, i} + v'_{\frac{\bp+\bq}{2}, i} \right) G^{\R}_{\bp} (\epsilon)
			\sigma^z G^{\A}_{\bq} (\epsilon)
		\right]
\label{eq:vartheta-1}
,\end{align}
where $\nu (\epsilon)$ is the density of states given by
\begin{align}
\nu (\epsilon)
	& = - \frac{2}{\pi N} \sum_{\bp} \Im [ G^{\R}_{\bp} (\epsilon) ]
	= \frac{1}{N} \sum_{\bp}
		F_{\bp} (\epsilon)
\end{align}
with the Cauchy distribution $F_{\bp} (\epsilon; \gamma)$ given by
\begin{align}
F_{\bp} (\epsilon; \gamma)
	& = \frac{1}{\pi} \frac{\gamma}{(\epsilon + \mu - T_{\bp} + \eta |\bm{\lambda}_{\bp}|)^2 + \gamma^2}
.\end{align}
It should be noted that Eq.~(\ref{eq:vartheta-1}) also gives the linear response of the charge current to $\dot{S}_z$. 
\begin{figure}[t]
\centering
\includegraphics[width=\linewidth]{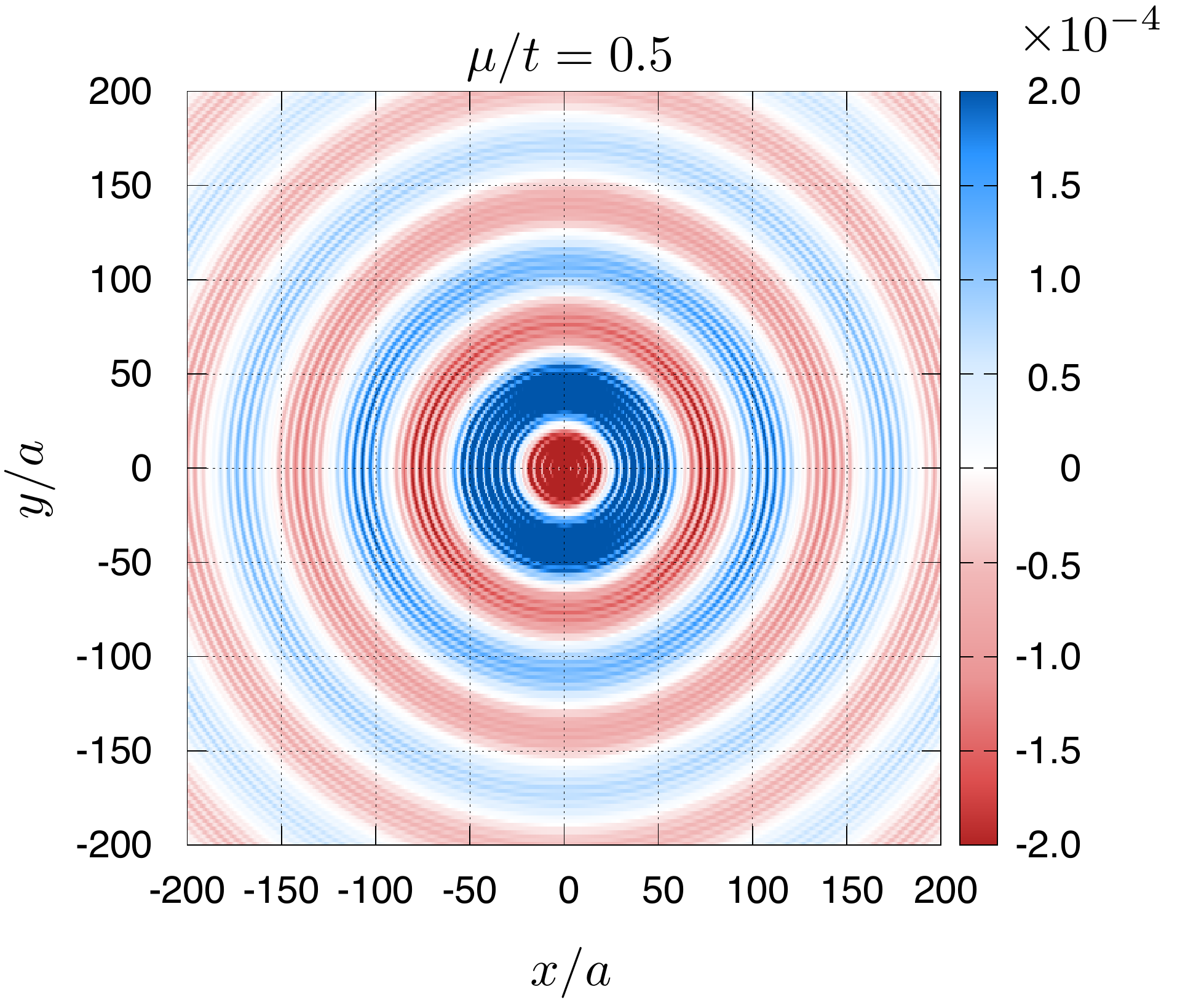}
\caption{\label{fig:D}The spatial profile of the charge current induced by the classical spin precession.
The unit of the current is $I_0 = e t / 8 \pi a \hbar$, the Rashba parameter $\tilde{\alpha}_{\R} = \alpha_{\R} / t a = 0.1$ and the exchange interaction strength $J_s / t = 1$.
The vortical structure is clearly shown. The clockwise current is represented by the red area and the counterclockwise current is described by the blue area. A more rapid oscillation is seen, which has the same origin as the Friedel oscillation~\cite{friedel1954} and RKKY interaction~\cite{ruderman1954,kasuya1956,yosida1957}.
}
\end{figure}
Taking the trace, we obtain
\begin{align}
\vartheta_i^{\alpha \beta} (\br)
	& = - \frac{\hbar J_s^2}{2} \epsilon^{\alpha \beta z}
		\int_{-\infty}^{\infty} \mathrm{d}\epsilon \left( - \frac{\partial f}{\partial \epsilon} \right) \nu (\epsilon)
\notag \\ & \hspace{-2em} \times
	\left[
	- \frac{8 \zi e a t}{\hbar} \left(
		- d_{i y} (\br, \epsilon) b_{i x} (\br, \epsilon)
		+ b_{i y} (\br, \epsilon) d_{i x} (\br, \epsilon)
	\right)
\right. \notag \\ & \hspace{-2em} \left.
	 - \frac{4 \zi e \alpha_{\R}}{\hbar} \epsilon^{z i \gamma} \epsilon^{\gamma z \delta}
	\left(
		- c_i (\br, \epsilon) b_{i \delta} (\br, \epsilon)
		- a_i (\br, \epsilon) d_{i \delta} (\br, \epsilon)
	\right)
	\right]
,\end{align}
where
\begin{align}
a_i (\br, \epsilon)
	& = - \frac{\pi}{2 N} \sum_{\eta = \pm} \sum_{\bp} e^{\zi \bp \cdot \br}
		\sin \frac{p_i a}{2}
		F_{\bp} (\epsilon; \gamma)
, \\
b_{i \alpha} (\br, \epsilon)
	& = - \frac{\pi}{2 N} \sum_{\eta = \pm} \sum_{\bp} e^{\zi \bp \cdot \br}
		\frac{\lambda^{\alpha}_{\bp}}{\epsilon + \mu - T_{\bp}}
		\cos \frac{p_i a}{2}
		F_{\bp} (\epsilon; \gamma)
, \\
c_i (\br, \epsilon)
	& = - \frac{\pi}{2 N} \sum_{\eta = \pm} \sum_{\bp} e^{\zi \bp \cdot \br}
		\cos \frac{p_i a}{2}
		F_{\bp} (\epsilon; \gamma)
, \\
d_{i \alpha} (\br, \epsilon)
	& = - \frac{\pi}{2 N} \sum_{\eta = \pm} \sum_{\bp} e^{\zi \bp \cdot \br}
		\frac{\lambda^{\alpha}_{\bp}}{\epsilon + \mu - T_{\bp}}
		\sin \frac{p_i a}{2}
		F_{\bp} (\epsilon; \gamma)
.\end{align}

Assuming that the chemical potential is low enough to approximate the dispersion as parabolic and setting $\mu + 4 t$ as $\mu$, we get the charge-current response
\begin{align}
\bm{J}^{\mathrm{dc}} (\br, t)
& = D (r) \bigl( \bm{S} (t) \times \dot{\bm{S}} (t) \bigr)^z \hat{e}_{\phi}
\label{eq:jdcT0}
,\end{align}
where $\hat{e}_{\phi} = (- \sin \phi, \cos \phi)$ with $\phi = \tan^{-1} y / x$ and the coefficient is given as $D (r) = D_1 (r) + D_2 (r)$ with
\begin{align}
D_1 (r)
	& = \frac{e J_s^2}{8}
		\int_{-\infty}^{\infty} \mathrm{d}\epsilon \left( - \frac{\partial f}{\partial \epsilon} \right) \nu (\epsilon)
	\frac{\hbar^2}{2 \me r} \left\{ \mathcal{B} (r, \epsilon) \right\}^2
, \\
D_2 (r)
	& = \frac{e J_s^2}{8}
		\int_{-\infty}^{\infty} \mathrm{d}\epsilon \left( - \frac{\partial f}{\partial \epsilon} \right) \nu (\epsilon)
	\alpha_{\R} \mathcal{C} (r, \epsilon) \mathcal{B} (r, \epsilon)
,\end{align}
and
\begin{align}
\mathcal{B} (r, \epsilon)
	& = \sum_{\eta} \int_0^{\infty} \mathrm{d}p p^2 J_1 (p r)
		\frac{\alpha_{\R}}{\epsilon + \mu - \frac{\hbar^2 p^2}{2 \me}}
		F_{p} (\epsilon; \gamma)
, \\
\mathcal{C} (r, \epsilon)
	& =  \sum_{\eta} \int_0^{\infty} \mathrm{d}p p J_0 (p r)
		F_{p} (\epsilon; \gamma)
.\end{align}
Here, $J_n (x)$ is the Bessel function of order $n$.

By taking the zero-temperature limit and the no-level-broadening limit, which means that $- \partial f / \partial \epsilon$ and $F_{\bp} (\epsilon;\gamma)$ are reduced to the delta functions $\delta (\epsilon)$ and $\delta (\epsilon + \mu - \hbar^2 p^2 / 2 \me + \eta \alpha_{\R} p)$, respectively, we get
\begin{align}
D_1 (r)
	& = \frac{e J_s^2 \me^2}{8 \hbar^4}
		\frac{\nu_{\mathrm{F}}}{p_0^2}
	\frac{\hbar^2}{2 \me r}
		\left( \sum_{\eta = \pm} \eta p_{\eta} J_{1} (|p_{\eta}| r) \right)^2
, \\
D_2 (r)
	& = - \alpha_{\R} \frac{e J_s^2 \me^2}{8 \hbar^4}
		\frac{\nu_{\mathrm{F}}}{p_0^2}
		\sum_{\eta' = \pm} |p_{\eta'}| J_{0} (|p_{\eta'}| r)
		\sum_{\eta = \pm} \eta p_{\eta} J_{1} (|p_{\eta}| r)
\end{align}
for $\mu \ge - \me \alpha_{\R}^2 / 2 \hbar^2 \equiv - E_{\R}$, and $D_1 (r) = D_2 (r) = 0$ for $\mu < - E_{\R}$.
In these expressions, $\nu_{\mathrm{F}} = \nu (0) = 1 / 2 \pi t$ for $\mu \ge 0$ and $\nu_{\mathrm{F}} = \me \alpha_{\R} / \hbar^2 p_0 t$ for $ - E_{\R} \le \mu < 0$ is the density of states at the Fermi level and 
\begin{align}
p_{\eta}
	& = p_0 + \eta \frac{\me}{\hbar^2} \alpha_{\R}
, \\
p_0
	& = \sqrt{ \frac{2 \me}{\hbar^2} \left( \mu + E_{\R} \right) }
.\end{align}

\begin{figure}[t]
\centering
\includegraphics[width=\linewidth]{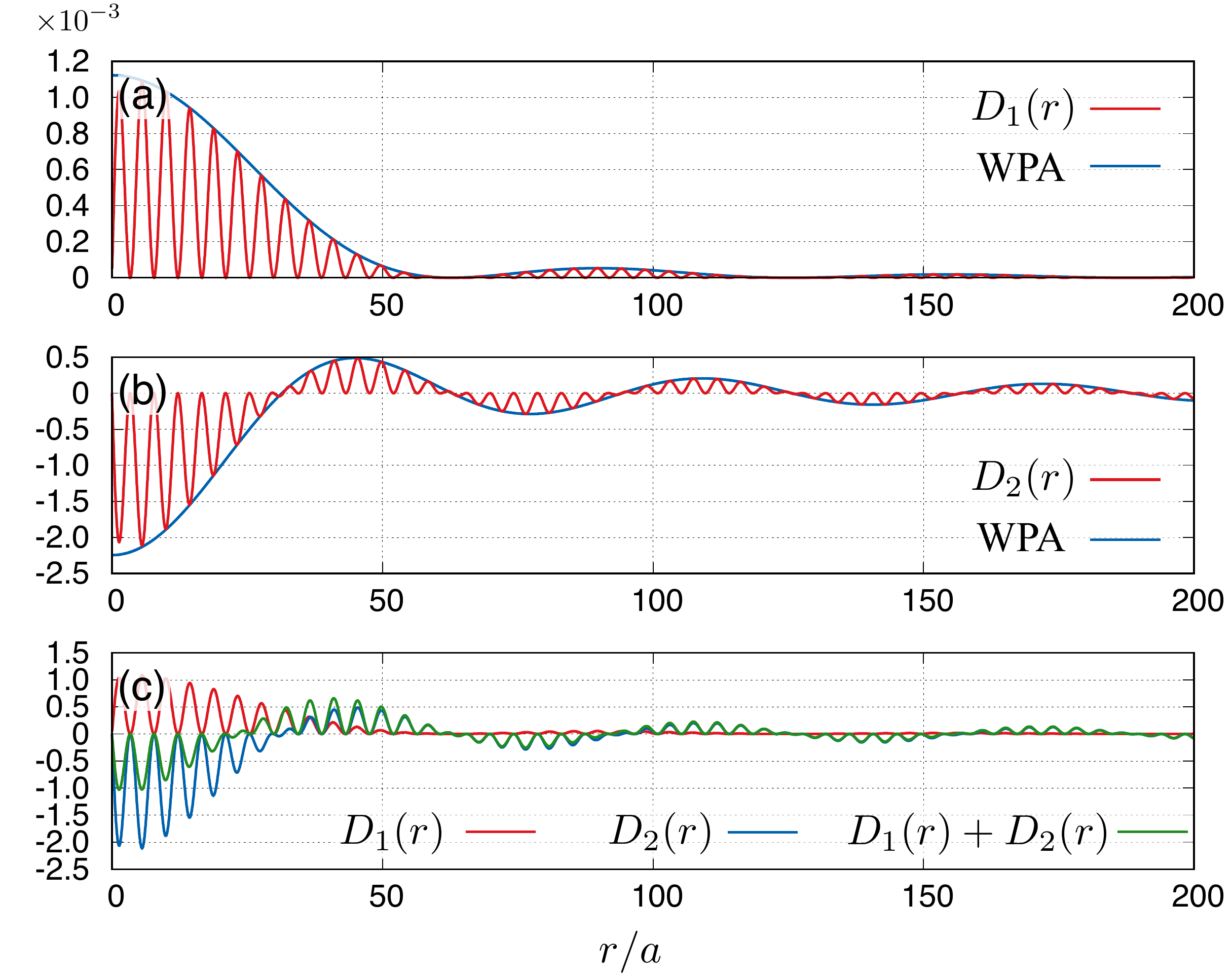}
\caption{\label{fig:D1-D2}(a) and (b):~The radial dependence of the response coefficients $D_1 (r)$ and $D_2 (r)$ at $T = 0$ for parameters $\mu/t=0.5$, $\tilde{\alpha}_{\R}=0.1$, $J_s / t = 1$, and $\gamma = 0$. The WPA results are also plotted. The envelopes of $D_1$ and $D_2$ are in good agreements with the WPA results. (c)~the radial dependence of total response coefficient.}
\end{figure}

The charge-current density given by Eq.~\eqref{eq:jdcT0} clearly forms a vortical structure.
Figure~\ref{fig:D} depicts the spatial profile of the current for Rashba parameter $\tilde{\alpha}_{\R} = \alpha_{\R} / t a = 0.1$ and chemical potential $\mu/t = 0.5$ with $\gamma = 0$. 
The changes between clockwise and counterclockwise current direction originate from the Rashba spin precession as shown by the wavepacket analysis~(WPA) in Ref.~\onlinecite{lange2021}. 
For more detail, we plot the radial dependence of the coefficients $D_1 (r)$, $D_2(r)$ and $D(r)$
in Fig.~\ref{fig:D1-D2}. 
We find that the envelopes of $D_1 (r)$ and $D_2 (r)$ agree with the corresponding WPA results [Eqs.~(9) and (10) of Ref.~\onlinecite{lange2021}]
\begin{align}
D_{1}^{\mathrm{WPA}} (x)
	& = A \frac{2 \sin^2 (\tilde{\alpha}_{\R} x / 2)}{\tilde{\alpha}_{\R} x^2}
\label{eq:D1_WPA}
, \\
D_{2}^{\mathrm{WPA}} (x)
	& = - A \frac{\sin (\tilde{\alpha}_{\R} x)}{x}
\label{eq:D2_WPA}
,\end{align}
where $x = r / a$, and $A = J_s^2 \tilde{\alpha}_{\R} / 2 \pi p_0 a t^2$.
The oscillations in $D_1 (r)$ and $D_2 (r)$ with shorter wavelength have the wavenumber $2 p_0 = p_{+} + p_{-}$, which comes from the interference of the two electron bands. They have the same origin as Friedel oscillations~\cite{friedel1954} and the Ruderman-Kittel-Kasuya-Yosida~(RKKY) interaction~\cite{ruderman1954,kasuya1956,yosida1957}.
\begin{figure}
\centering
\includegraphics[width=\linewidth]{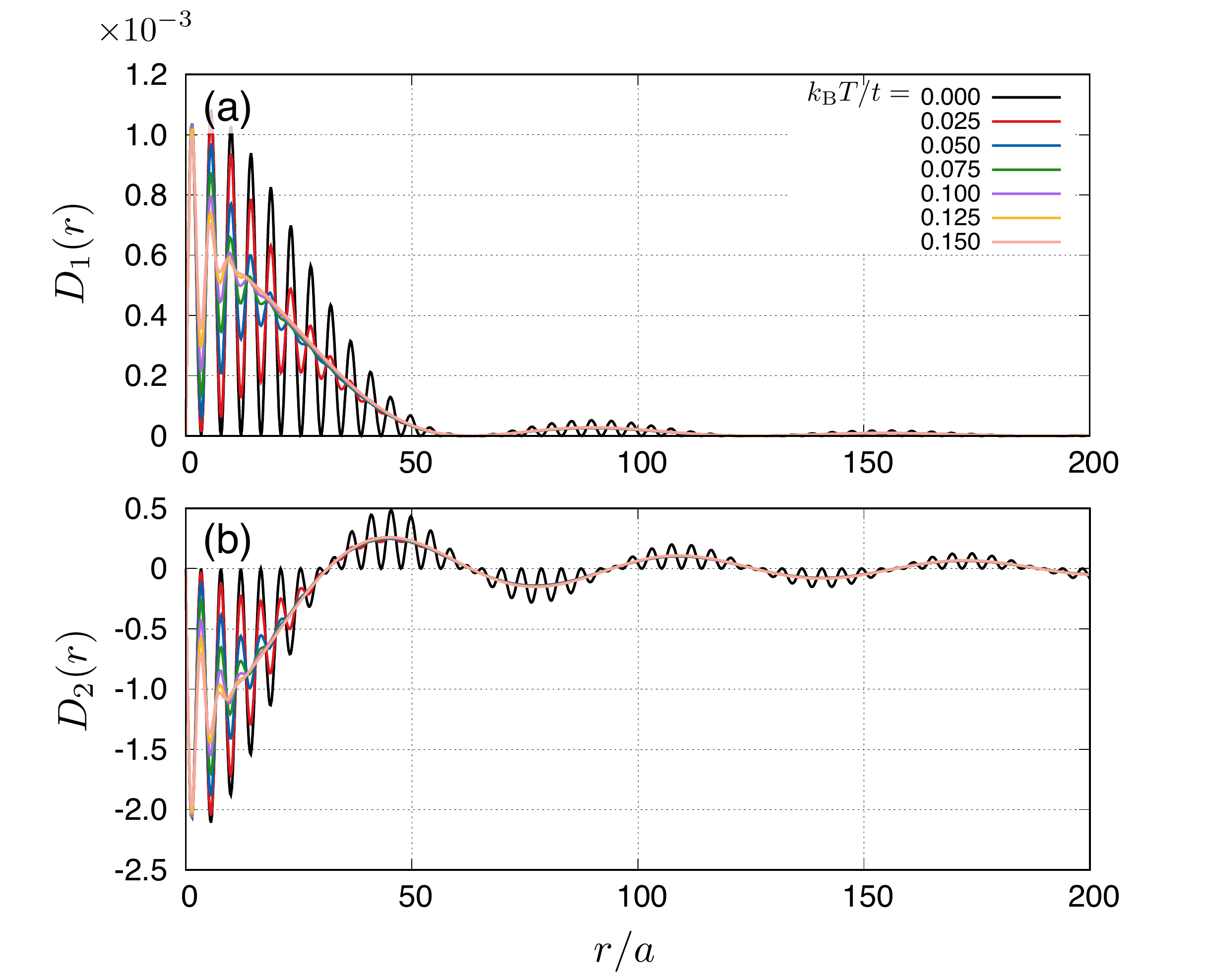}
\caption{\label{fig:finiteT}
  The current profile at finite temperatures.
  We set level-broadening parameter $\gamma / t = 5 \times 10^{-4}$ and the other model parameters are as in Figs.~\ref{fig:D} and \ref{fig:D1-D2}. 
}
\end{figure}
Figure~\ref{fig:finiteT} shows the temperature dependence of the current. 
For finite $T$, the rapid oscillations become weaker as the radius increases and eventually disappear.  

We can see from Eqs.~(\ref{eq:D1_WPA}) and (\ref{eq:D2_WPA}) that the oscillation period is described by the Rashba parameter.
This dependence allows us to estimate the strength of the Rashba SOC.
The Rashba SOC amplitude may also be determined from the magnetic field induced by the current vortex~\cite{lange2021}.

\begin{figure}
\centering
\includegraphics[width=0.93\linewidth]{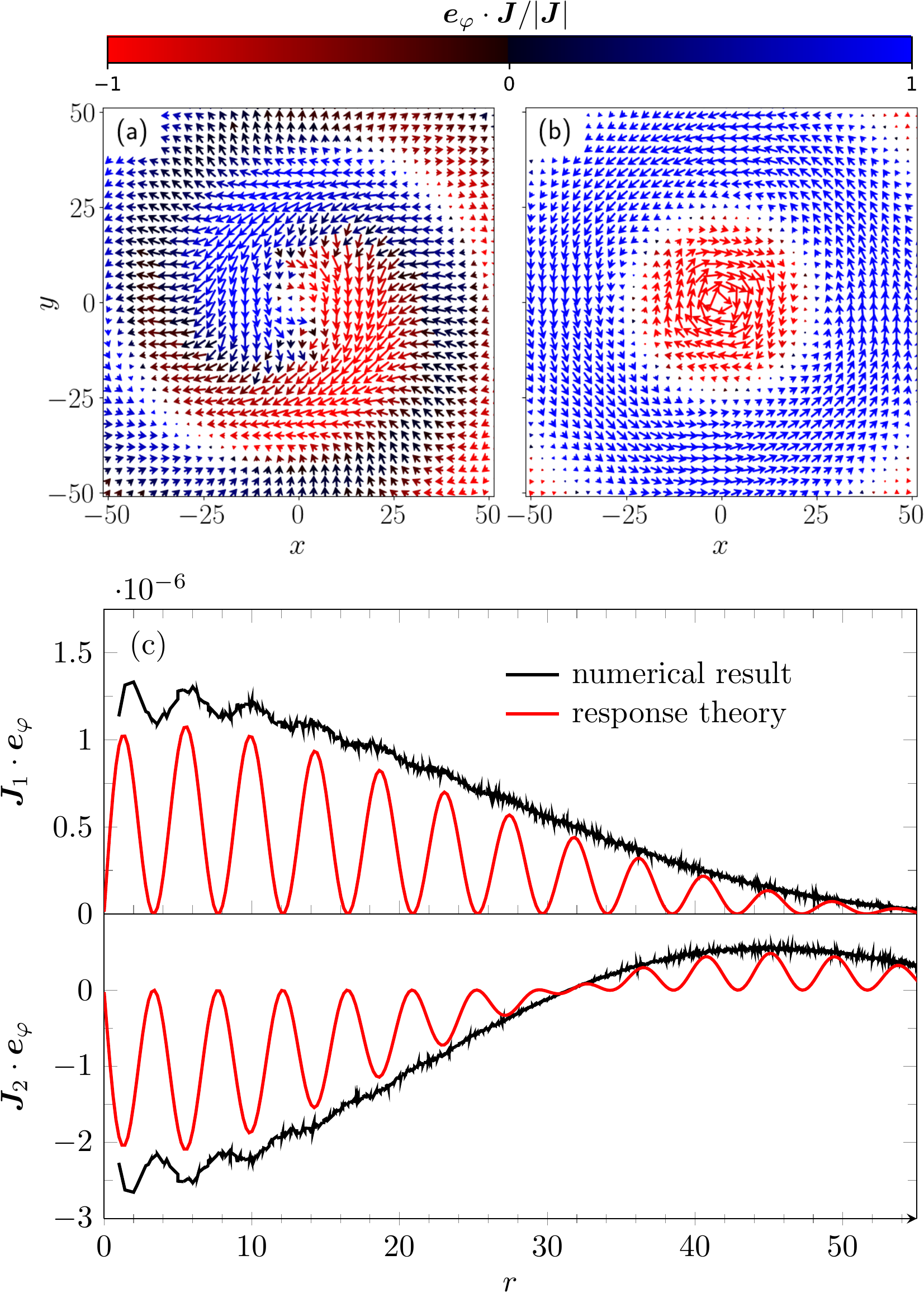}
\caption{\label{fig:numerical}Numerically calculated charge current due to spin pumping at the origin at $T=0$. Model parameters are $J_s/t=0.1$, $\omega =0.1 t/\hbar$, $\mu/t=0.5$ and $\tilde{\alpha}_{\R}=0.1$. The color of the arrows indicates the charge-current direction, their lengths the magnitude. (a):~snapshot at time $100\,\hbar/ t$. (b):~time average over a period $1/\omega$ around $100\,\hbar/ t$. The scale in (a) is larger than in (b) by a factor 20. (c):~radial dependence of the current in (b) compared with response theory. The currents $\bm{J}_1$ and $\bm{J}_2$ correspond to the response coefficients $D_1(r)$ and $D_2(r)$, respectively. As before, the current is given in units of $I_0 = e t / 8 \pi a \hbar$.}
\end{figure}

To check the response-theory results, we performed numerical calculations. 
We use a Lanczos transformation to map the Hamiltonian to a chain representation, and employ the infinite boundary conditions described in Ref.~\onlinecite{PhysRevB.88.245113}.
The time-dependent currents are then obtained by a straightforward numerical diagonalization of the one-electron Hamiltonian.
Details of the numerical method can be found in the Suplemental Material of Ref.~\onlinecite{lange2021}. 
Figure~\ref{fig:numerical} displays results for an exchange-coupling strength $J_s / t = 0.1$ and precession frequency $\omega = 0.1 t/\hbar$ at $T=0$. 
The charge current at some fixed time is not symmetric under rotation around the origin and deviates strongly from the analytical expression~\eqref{eq:jdcT0} for the dc component [Fig.~\ref{fig:numerical} (a)].
After averaging over a period $1/\omega$, however, we recover the vortical structure predicted by the response theory [Fig.~\ref{fig:numerical} (b)]. 
While oscillations with wavenumber $2 p_0$ occur at small $r$, they are not nearly as strong as in the analytical result. This may be due to the finite frequency $\omega$, which could have a similar effect as a finite temperature by inducing excitations that are not exactly at the Fermi energy. 

We would like to make a comment concerning the connection to our previous work.
In Ref.~\onlinecite{lange2021}, we considered a Rashba system coupled to a quantum spin chain. The spin injection was achieved by driving a spinon spin current through the spin chain. In this paper, we instead investigate the spin injection by local spin pumping.
Both ways of the spin injection lead to the same result of the current vortex generation in the Rashba system.
Hence, we conclude that it is a universal phenomenon that the local spin injection induces the current-vortex generation in the Rashba system.
It should be noted that we discuss the temperature dependence of the curren-vortex generation in the present work.

We also comment on related materials for studying the current-vortex generation.
Since we consider a Rashba system which is coupled to a classical spin, 2D oxides~\cite{benshalom2010,king2012}, Ag/Bi surfaces~\cite{ast2007}, and some van der Waals materials~\cite{wang2017a,zhang2018a} are examples of our model, if a nano-scale magnet could be attached.

Moreover, we give a comment concerning the inverse effect of current-vortex generation by local spin injection, i.e., local spin extraction induced by a charge current vortex in the Rashba system.
In the Rashba system, charge-to-spin conversion occurs because of the Edelstein effect~\cite{edelstein1990}.
Although such charge-to-spin conversion is interesting, it would be difficult to prepare the current vortex around a nano magnet.

\section{\label{sec:summary}Summary}
Spin-orbit coupling phenomena, such as the (inverse) spin Hall effect, are of central importance in spintronics, as they facilitate the conversion between spin and charge degrees of freedom. 
A minimial model to investigate these effects theoretically is the free two-dimensional electron gas with Rashba spin-orbit coupling. 
Mesoscopic systems and interface effects are more easily studied using the lattice version of the Rashba model, however. In this paper, we demonstrated how the spin- and charge-current operators on a lattice and in the presence of spin-orbit coupling can be derived from appropriate gauge transformations. 
We furthermore showed that, like in the continuum model, the spin-orbit coupling Hamiltonian can be defined as the generator of the intrinsic spin current. 

As a specific application of the lattice description, we then analyzed the charge-current response of the Rashba model to spin injection via local spin pumping which we modeled by coupling a single site of the lattice to a classical precessing spin. 
The dc component of the charge current was found to form a vortical structure, where the current direction depends on the distance from the classical spin. 
This dependence is explained by the Rashba spin precession. 
We also found additional spatial oscillations with twice the Fermi wavenumber, which have a similar origin as Friedel oscillations and the RKKY interaction.
At finite temperature, these oscillations disappear for sufficiently long distances. 
Lastly, we performed numerical simulations, which include effects beyond the contributions considered in the analytic calculation. 
Although the snapshot of the spatial profile of the current in a certain time has a complicated structure, the time average is consistent with the dc component from the analytic calculation.

While we focused here on the Rashba system, the local spin injection will generally induce the nonuniform charge currents in spin-orbit coupled systems, where the spatial profile of the current should reflect the electronic structure of the system.
Hence, the spin-charge conversion by local spin injection has a great potential to investigate the electronic structure such as the SOC strength. 

The data that support the findings of this study are available from the corresponding author upon reasonable request.
JF is partially supported by the Priority Program of Chinese Academy of Sciences, Grant No. XDB28000000.
SE and FL are supported by Deutsche Forschungsgemeinschaft through project EJ7/2-1 and FE398/8-1, respectively.
SM is supported by JST CREST Grant (nos. JPMJCR19J4, JPMJCR1874 and JPMJCR20C1) and JSPS KAKENHI (nos. 17H02927 and 20H01865) from MEXT, Japan.
JF thanks Y.~Ominato for valuable discussion.

 \appendix
\section{\label{apx:continuity}Continuity equations}
In this section, we derive the charge and spin  continuity equations for the tight-binding Hamiltonian $\mathcal{H} = \mathcal{H}_t + \mathcal{H}_{\R}$ with Rashba SOC. 
The continuity equation of the charge $Q_{n,m} = - e c^{\dagger}_{n,m} c^{}_{n,m}$
is evaluated from
\begin{align}
\dot{Q}_{n,m}
	& = - e \left(
		  \dot{c}^{\dagger}_{n,m} c^{}_{n,m}
		+ c^{\dagger}_{n,m} \dot{c}^{}_{n,m}
	\right)
\label{eq:continuity_eq_charge}
\end{align}
with the time derivative of the field operators $c_{n,m}$ given by the Heisenberg equation of motion
\begin{align}
\dot{c}_{n,m}
	& = \frac{1}{\zi \hbar} \left[ c_{n,m}, \mathcal{H} \right] \nonumber \\ 
	& = - \frac{t}{\zi \hbar} \left(
		  c^{}_{n+1,m}
		+ c^{}_{n-1,m}
		+ c^{}_{n,m+1}
		+ c^{}_{n,m-1}
	\right)
\notag \\ & \hspace{-1em}
	- \frac{\alpha_{\R}}{2 \hbar} \left(
		\sigma^{y} \frac{ c^{}_{n+1,m} - c^{}_{n-1,m} }{a}
		- \sigma^{x} \frac{ c^{}_{n,m+1} - c^{}_{n,m-1} }{a}
	\right)
\label{eq:dot_c}
.\end{align}
Substituting Eq.~(\ref{eq:dot_c}) 
into Eq.~(\ref{eq:continuity_eq_charge}) results in 
\begin{align}
\dot{Q}_{n,m}
	& = \Gamma_{t, x} + \Gamma_{t, y} + \Gamma_{\R, x} + \Gamma_{\R, y}
,\end{align}
where we have suppressed the indices $n$ and $m$ on the right-hand side, and
\begin{align}
\Gamma_{t, x}
	& = \frac{\zi e t}{\hbar} \Bigl\{
		\left(
			  c^{\dagger}_{n+1,m}
			+ c^{\dagger}_{n-1,m}
		\right) c_{n, m}
		- \mathrm{h.c.}
	\Bigr\}
, \\
\Gamma_{t, y}
	& = \frac{\zi e t}{\hbar} \Bigl\{
		\left(
			  c^{\dagger}_{n,m+1}
			+ c^{\dagger}_{n,m-1}
		\right) c_{n, m}
		- \mathrm{h.c.}
	\Bigr\}
, \\
\Gamma_{\R, x}
	& = \frac{e \alpha_{\R}}{2 \hbar} \left(
		\frac{ c^{\dagger}_{n+1,m} - c^{\dagger}_{n-1,m} }{a} \sigma^{y} c_{n,m}
		+ \mathrm{h.c.}
	\right)
, \\
\Gamma_{\R, y}
	& = - \frac{e \alpha_{\R}}{2 \hbar} \left(
		\frac{ c^{\dagger}_{n,m+1} - c^{\dagger}_{n,m-1} }{a} \sigma^{x} c_{n,m}
		+ \mathrm{h.c.}
	\right)
.\end{align}
In the continuum limit, the first term becomes
\begin{align}
\Gamma_{t, x}
	& = \frac{\zi e t a^2}{\hbar} \left(
		\frac{ c^{\dagger}_{n+1,m} - 2 c^{\dagger}_{n,m} + c^{\dagger}_{n-1,m} }{a^2} c_{n, m}
		- \mathrm{h.c.}
	\right)
\notag \\
	& \to \frac{\zi e \hbar}{2 \me} \left(
		\frac{\partial^2 c^{\dagger}}{\partial x^2} c
		- c^{\dagger} \frac{\partial^2 c}{\partial x^2}
	\right)
\notag \\
	& = - \frac{\partial j_{x} (\br)}{\partial x}
,\end{align}
where we have used 
Eq.~(\ref{eq:j-continuum}).
Similarly, 
$\Gamma_{t, y} \to \partial j_y (\br) / \partial y$, so that
\begin{align}
\Gamma_{t, x} + \Gamma_{t, y}
	& \to - \bm{\nabla} \cdot \bm{j}
.\end{align}
The continuum limits of the Rashba terms are [see Eq.~\eqref{eq:j-continuum_R}] $\Gamma_{\R, x} \to (e \alpha_{\R} / \hbar) \partial (c^{\dagger} \sigma^y c) / \partial x$ and $\Gamma_{\R, y} \to - (e \alpha_{\R} / \hbar) \partial (c^{\dagger} \sigma^x c) / \partial y$, which may be summarized to 
\begin{align}
\Gamma_{\R, x} + \Gamma_{\R, y}
	& \to - \bm{\nabla} \cdot \bm{j}'
.\end{align}
From the above, we finally obtain the continuity equation
\begin{align}
  \dot{Q} = - \bm{\nabla} \cdot (\bm{j} + \bm{j}')
\end{align}
that expresses the conservation of charge.

We now derive the continuity equation for spin in the same manner. The time-derivative of the spin operators is
\begin{align}
\dot{s}^{\alpha}_{n,m}
	& = \frac{\hbar}{2} \left(
		  \dot{c}^{\dagger}_{n,m} \sigma^{\alpha} c^{}_{n,m}
		+ c^{\dagger}_{n,m} \sigma^{\alpha} \dot{c}^{}_{n,m}
	\right)
\label{eq:continuity_eq_spin}
.\end{align}
Substituting Eq.~(\ref{eq:dot_c}) into this expression, we have
\begin{align}
\dot{s}^{\alpha}_{n,m}
	& = \Sigma^{\alpha}_{t, x}
		+ \Sigma^{\alpha}_{t, y}
		+ \Sigma^{\alpha}_{\R, x}
		+ \Sigma^{\alpha}_{\R, y}
,\end{align}
with
\begin{align}
\Sigma^{\alpha}_{t, x}
	& = \frac{t}{2 \zi} \Bigl\{ \left(
		  c^{\dagger}_{n+1,m}
		+ c^{\dagger}_{n-1,m}
	\right) \sigma^{\alpha} c_{n,m}
	- \mathrm{h.c.}
	\Bigr\}
, \\
\Sigma^{\alpha}_{t, y}
	& = \frac{t}{2 \zi} \Bigl\{ \left(
		  c^{\dagger}_{n,m+1}
		+ c^{\dagger}_{n,m-1}
	\right) \sigma^{\alpha} c_{n,m}
	- \mathrm{h.c.}
	\Bigr\}
, \\
\Sigma^{\alpha}_{\R, x}
	& = - \frac{\alpha_{\R}}{4} \left(
		\frac{ c^{\dagger}_{n+1,m} - c^{\dagger}_{n-1,m} }{a} \sigma^{y} \sigma^{\alpha} c_{n,m}
	+ \mathrm{h.c.}
	\right)
, \\
\Sigma^{\alpha}_{\R, y}
	& = \frac{\alpha_{\R}}{4} \left(
		\frac{ c^{\dagger}_{n,m+1} - c^{\dagger}_{n,m-1} }{a} \sigma^{x} \sigma^{\alpha} c_{n,m}
	+ \mathrm{h.c.}
	\right)
        ,\end{align}
where the indices $n$ and $m$ on the right-hand side are again suppressed. 
In the continuum limit,
\begin{align}
\Sigma^{\alpha}_{t, x}
	& = \frac{t a^2}{2 \zi} \left(
		\frac{ c^{\dagger}_{n+1,m} - 2 c^{\dagger}_{n,m} + c^{\dagger}_{n-1,m} }{a^2} c_{n, m}
		- \mathrm{h.c.}
	\right)
\notag \\
	& \to \frac{\hbar}{2} \frac{\hbar}{2 \me \zi} \left(
		\frac{\partial^2 c^{\dagger}}{\partial x^2} \sigma^{\alpha} c
		- c^{\dagger} \sigma^{\alpha} \frac{\partial^2 c}{\partial x^2}
	\right)
\notag \\
	& = - \frac{\partial j_{\mathrm{s}, x}^{\alpha} (\br)}{\partial x}
,\end{align}
and similarly $\Sigma^{\alpha}_{t, y} \to \partial j_{\mathrm{s}, y}^{\alpha} (\br) / \partial y$, so that
\begin{align}
\Sigma^{\alpha}_{t, x} + \Sigma^{\alpha}_{t, y}
	& \to - \bm{\nabla} \cdot \bm{j}_{\mathrm{s}}^{\alpha} (\br)
.\end{align}
For the Rashba part, we obtain 
\begin{align}
\Sigma^{\alpha}_{\R, x} + \Sigma^{\alpha}_{\R, y}
	& \to - \bm{\nabla} \cdot \bm{j}_{\mathrm{s}}^{(1), \alpha}
		+ \tau^{\alpha}
\end{align}
with the torque term
\begin{align}
\tau^{\alpha}
	& = \frac{\alpha_{\R}}{t a^2} \sum_{i,j,\beta} \epsilon^{i j z} \epsilon^{\alpha j \beta} j_{\mathrm{s}, i}^{\beta} (\br)
.\end{align}
Therefore, the continuity equation for spin in the continuum limit is
\begin{align}
\dot{s}^{\alpha}
	& = - \bm{\nabla} \cdot \left( \bm{j}_{\mathrm{s}}^{\alpha} + \bm{j}_{\mathrm{s}}^{\prime \alpha} \right) + \tau^{\alpha}
\label{eq:continuity_eq_s_continuum}
.\end{align}
As indicated by the torque term $\tau^{\alpha}$, the spin is not conserved. The underlying reason is that the Rashba SOC breaks the $\mathrm{SU}(2)$ rotational symmetry in spin space.

It should be noted that the spin current for the lattice Rashba Hamiltonian is defined by Eqs.~(\ref{def1:jsx_R}) and (\ref{def1:jsy_R}), in which the additional factor $1/2$ is included. The continuity equation~(\ref{eq:continuity_eq_s_continuum}) holds for the spin current $\bm{j}_{\mathrm{s}}^{\prime \alpha}$ as given by Eq.~(\ref{eq:js_R_continuum}).

\bibliography{reference,reference_f}
\end{document}